\documentclass[]{article}
\usepackage[utf8]{inputenc}
\usepackage[english]{babel}

\usepackage{amsmath}
\usepackage{amsfonts}
\usepackage{graphicx}
\usepackage[left=5cm,right=5cm,top=2cm,bottom=2cm]{geometry}

\usepackage{nicefrac}       % Cool fractions  (⌐■_■)

\newcommand{\Fig}[1]{Fig.{$\;$}(#1)}         % Figures
\newcommand{\coma }{,}                       % Coma for the equations
\newcommand{\punto}{.}                       % Punto for the equations

\usepackage[colorlinks=true,linkcolor=blue,urlcolor=blue]{hyperref}

\title{A lite parametric model for the Hemodynamic Response Function}

\author{M. Morante}

\begin{document} %#######################################################################
\maketitle

\begin{abstract}
	When working with task-related fMRI data, one of the most crucial parts of the data analysis consists of determining a proper estimate of the BOLD response. The following document present a lite model for the Hemodynamic Response Function HRF. Between other advances, the proposed model present less number of parameters compare to other similar HRF alternative, which reduces its optimization complexity and facilitates its potential applications.
	
\end{abstract}
\section{Introduction}
Functional Magnetic Resonance Imaging (fMRI) is one of the most dominant data acquisition technique for the detection and study of brain activity. However, the detection of the brain activity is indirect: fMRI measures the blood oxygenation level-dependent (BOLD) contrast~\cite{Handbook,FunMRI}, which constitutes the evoked hemodynamic response of the brain to the corresponding neuronal activity.

In the case of task-related fMRI experimental designs, the participants are presented with a fixed number of pre-selected stimulus. One way to proceed with the analysis of these fMRI data is to estimate the shape of the time courses that correspond to the considered stimulus of the experimental design. There exits several alternatives to estimate these time courses. The most common procedure is the linear convolutional model, which assumes that the task-related time courses can be modeled as a convolution between the actual neuronal activation responses and a particular impulse response function referred to as Hemodynamic Respones Function (HRF).

Conventional fMRI data analysis methods require an explicit estimate of the functional shape of the HRF to infer the hemodynamic response of the measured BOLD sequences. For example, for those based on the General Linear Model (GLM), the accurate estimate of the HRF is crucial~\cite{FunMRI}. The most widely used model for the functional shape of the HRF is the double gamma distribution model~\cite{FristonOri}, usually referred to as the canonical HRF, which in one of the default HRF function for different software packages such as SPM~ \cite{SPM-web,SPM-Book}. Other alternatives such cosine function~\cite{Cosine_2002}, radial bases~\cite{Riera_2004}, spectral basis functions~\cite{Liao_2002} are potential alternatives. However, note that not all the models are equally good for capturing the evoked changes of the HRF neither presents the same number of parameters \cite{HRF-models}.

%On the other hand, advances in data fusion, in the sense of analyzing several dataset in a way that they can interact and share information between eachother, have bring together fMRI and EEG. From a general perspective, fMRI and EEG dataset share important information regarding different parts of the brain behavior and combined information may help to enhance our knowledge regarding the brain. Nonetheless, most of these approaches still implement the convolution linear model to relay the information between modalities. In this way, a propper functional shape of the HRF is required in order to properly model the brain-responses.

However, the convolution model presents several limitations. First, it assumes that the HRF is known and fixed. No doubt, this is a strong assumption and controversial from a practical point of view, since the truth is that the hemodynamical response --especially its latency-- varies across different brain regions~\cite{HRF-Var} as well as between different subjects~\cite{HRF-Var-2}.

On the other hand, other multiple factors may affect the functional shape of the hemodinamic response. For example, short interstimulus interval may introduce non-linear effects that may compromise the linearity of the convolutional model. In this way, non-lineal model have beel also proposed that accurately encapsulate the subjects variability. However, these model often require an extensive number of parameters, which may obscure its application in practice.

For all these reason, an new parametric model for the functional shape of the HRF model is presented in this work. The proposed model requires a relatively low number of parameters and present a more suitable optimization properties compared to similar alternatives.

\section{Modeling of the BOLD response}
In general, the evoked BOLD response in fMRI is a complex and non-linear function of the neuronal and vascular changes induced by the neuronal activity~\cite{Accounting_2005}. Thus, the shape of the response depends both on the applied stimulus and the hemodynamic response to the neuronal events.

As we briefly mention above, nowadays there exists several alternative to model this behavior. Non-linear models seems to be a natural assumption, since they encapsulates the real nature of the BOLD response. However, linear models often provide more robust and interpretable results. In this way, linear models have been widely used for modeling the BOLD response, where linearity implies that the magnitude and shape of the evoked HRF do not depend on any preciding stimuli. Despite the potential limitations, studies have shown that linear models works well under certain conditions, particularly when the events are sufficiently spaced in time. 

Within the lineal framework, the convolutional model is one of the most popular models, which assumes that the measured signal $x(t)$, at the time instant $t$, is obtained as the convolution of a stimulus function $u(t)$ and the HRF $h(t)$, that is,
\begin{equation}
	x(t) = u(t)*h(t)
	\punto
\end{equation}

Regarding the HRF, from observing the natural behavior of the hemodynamic response~\cite{FunMRI}, there are two main phases that drive the functional shape of the HRF. First, after the neuronal activation starts, the surrounding tissues reacts triggering several metabolic reactions, in particular, increasing the oxygen consumption which alter the magnetic properties of the environment, which is what causes the characteristic peak of the HRF. Besides, the vascular system reacts expanding the blood vessel to deploy oxygen. Then, after the neuronal activity stops, the oxygen consumption drops but the vascular system continues deploying oxygen until the blood vessels relax to the normal state. This over-compensatory stage produces the characteristic ``undershoot'' of the HRF.

\begin{figure}[h]
	\hspace{-2cm}
	\includegraphics[scale=0.5]{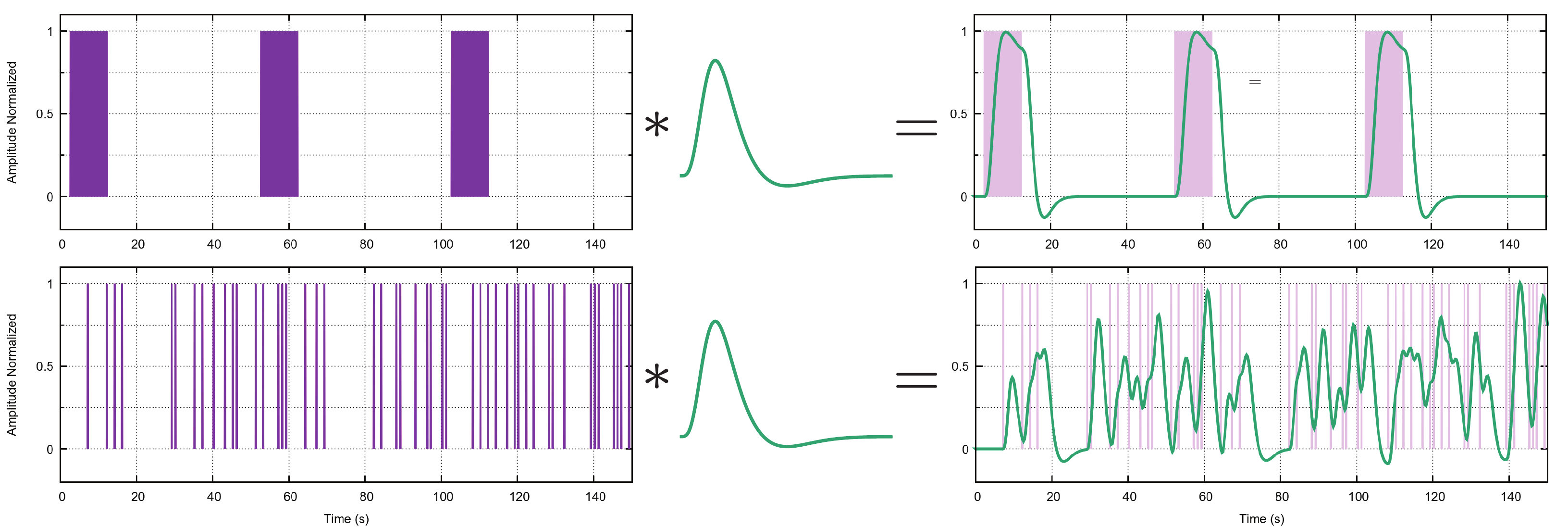}
	\caption{Convolution Model. This figure shows an example of two time courses using the convolutional model. The first experimental design correspond to a block-related example and the second is an example of an event-related condition.}
	\label{Fig:ConvMod}
\end{figure}

On the other hand, several studies have shown evidence of a decrease in oxygenation levels in the time inmediately following the neuronal activity, given an small decrease in the BOLD signal~\cite{Lindquist_2008}. However, the existence of this initial deep has not completely confirmed and its existence remains controversial~\cite{Can_2000}.

In this way, using the a proper estimate of the HRF and the empirical design, $u(t)$, we can obtain an estimate of the time courses. \Fig{\ref{Fig:ConvMod}} shows an example of the response obtained using the same HRF with different stimulus.

\subsection{The canonical HRF}
There exists several models to describe the functional shape of the HRF. As we mentioned above, one of the most widely used is the two-gamma distribution model~\cite{HRF-Var-2}, usually referred to as the canonical HRF. In general, the canonical HRF in a parametric model that can be written as:
\begin{equation}
	h(t)=\frac{t^{a_{1}-1}e^{-a_{2}t}}{\Gamma(a_{1})}-\alpha\frac{t^{a_{3}-1}e^{-a_{4}t}}{\Gamma(a_{3})}
	\coma
\end{equation}
where $\Gamma(\cdot)$ is the Gamma function, $alpha$ controls the relative rati of response to undershoot, and $a_{1},a_{2},a_{3}$, and $a_{4}$ are four parameters that controls the functional shape of the HRF. Therefore, for this model, a total number of 5 parameters are required to define the HRF.

\begin{figure}[h]
	\centering
	\includegraphics[scale=0.7]{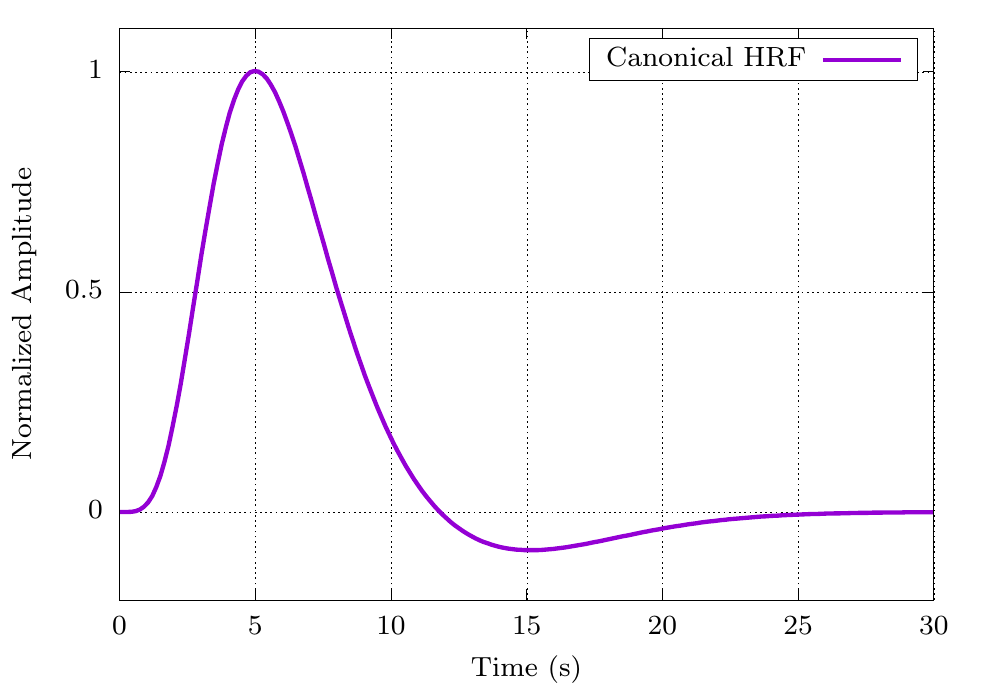}
	\caption{Graphical representation of the canonical HRF.}
	\label{Fig:canHRF}
\end{figure}

For this model, \Fig{\ref{Fig:canHRF}} show the HRF corresponding to the parameter the values $\alpha=\frac{1}{6}$, $a_{1}=6$, $a_{3}=16$ and $a_{2}=a_{4}=1$, which are the parameters usually selected for the definition of the canonical HRF~\cite{SPM-web}.

\section{Novel lite model for the HRF}
According to the natural behavior of the thermodynamic response, a good model should ideally take in consideration, first, the increasing on oxygen consumption and vascular coupling and, second, the over-compensatory oxygen effect and the vascular relaxation. In this way, a good model should ideally take in cosideration these two phenomena with the less number of parameters as possible.

For example, the repulsive and attractive force that neutral atoms or molecules is  often model with a very simple mathematical model referred to Leonard-Jones potential (6,12). Despite its simplicity and relatively low number of parameters, the model works well and allows to produce good estimates.

In this way, and based on the previous models, we propose a new Leonard-Jones Motivated (LJM) parametric model for the functional shape of he HRF it can be written as:
\begin{equation}
	h(t)=\Gamma^{-3}(at)-\alpha\Gamma^{-6}(bt)
	\coma
\end{equation}
where $\Gamma (\cdot)$ is the Gamma function, $a$ controls the main shape and position of the maximum of the HRF maximum and $\alpha,b$ controls the shape and relative position of the HRF's undershoot. Unlike the cannonical HRF, the proposed model present only 3 parameters, which eases its application in practice.

Similar to the Leonard Jones potential, the proposed model encapsulates the two major physical response of the brain activity with a reduced number of parameters. Besides, in this model we particularly selected the indices (3,6) because they were the parameters that better fits the functional shape of the HRF. However, is the same way as the LJ potential allows, these indices can be relaxed or even tunned as extra parameters if necessary. The behavior of these indices will also explore in future works.

Finally, for this model, \Fig{tal} shows the HRF corresponding to the parameters blablabla, compared with the canonical HRF. Note that the HRF are similar but not the same as we expected, since they are different models.

\begin{figure}[h]
	\centering
	\includegraphics[scale=0.7]{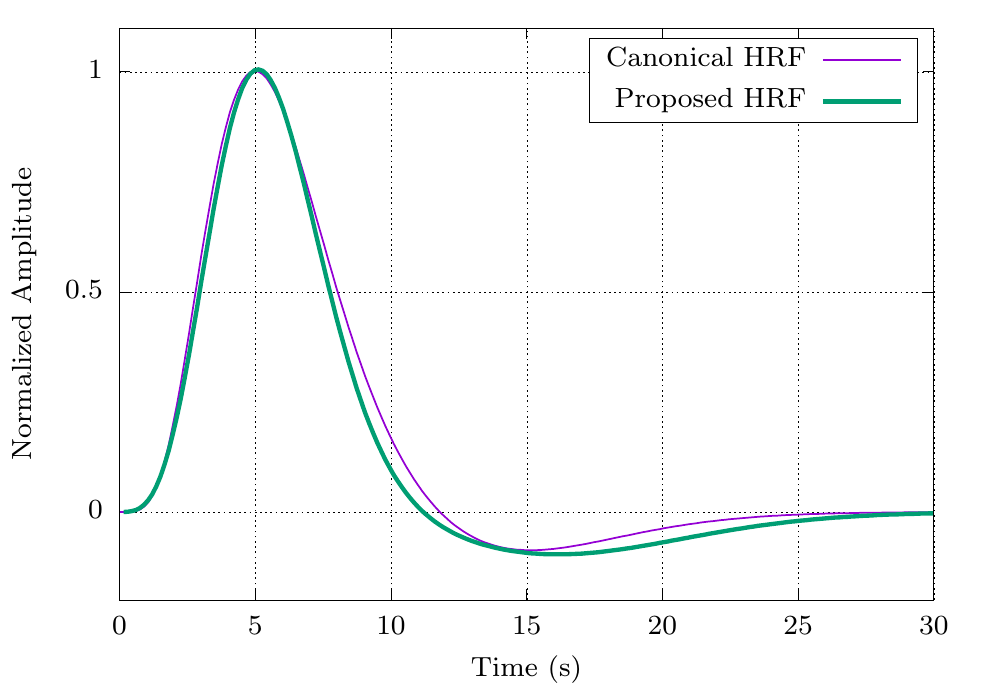}
	\caption{Representation of the proposed HRF (green) compared to the standard canonical HRF (purple).}
\end{figure}

\subsection*{Derivatives}
The proposed model is fully differentiable. Here we present some relevant derivatives:
\paragraph{First derivative}
The first derivative it can be obtained as:
\[
	h^{\prime}(t)=\frac{\partial h}{\partial t}=-3a\Gamma^{-4}(at)\Gamma^{\prime}(at)+6\alpha b\Gamma^{-7}(bt)\Gamma^{\prime}(bt)
	\coma
\]
where using the properties of the Gamma function, the derivative can be written as:
\begin{equation}
	h^{\prime}(t)=-3a\psi_{0}(at)\Gamma^{-3}(at)+6\alpha b\psi_{0}(bt)\Gamma^{-6}(bt)
\end{equation}
where $\psi_{0}$ is the polygamma function of order $0$, also called the polygamma function.

\paragraph{Second derivative}
Similarly, the second derivative can be written as:
\begin{equation}
	\begin{aligned}
	h^{\prime\prime}(t)= \frac{\partial^{2}h}{\partial t^{2}} = & 3a^{2}\left(3\psi_{0}^{2}(at)-\psi_{1}(at)\right)\Gamma^{-3}(at)- \\ 
	& -6\alpha b^{2}\left(6\psi_{0}^{2}(bt)-\psi_{1}(bt)\right)\Gamma^{-6}(bt)
	\end{aligned}
	\coma
\end{equation}
where $\psi_{1}$ is the polygamma function of order 1.

\subsubsection*{Partial derivatives with respect the parameters}
Concerning the behavior of the proposed model with respect to the three parameters, the following partial derivative can be obtained:

\paragraph{Parameter $\alpha$:}
\begin{equation}
	\frac{\partial h}{\partial\alpha}=-\Gamma^{-6}(bt)
	\coma
\end{equation}
which has a single local minimum at the point $t=\frac{x_0}{b}$, where $x_{0}=1.461632144\ldots$, that is, the first zero of the digamma function, $\psi_{0}$.

\paragraph*{Parameter a:}
\begin{equation}
	\frac{\partial h}{\partial a}=-3t\psi_{0}(at)\Gamma^{-3}(at)
	\punto
\end{equation}

\paragraph{Parameter b:}
\begin{equation}
	\frac{\partial h}{\partial b}=-6\alpha bt\psi_{0}(bt)\Gamma^{-6}(bt)
	\punto
\end{equation}

\paragraph{Derivatives relation:}
Interestingly, we can observe that there exists an particular relation between the different partial derivatives:
\begin{equation}
	t\frac{\partial h}{\partial t}=a\frac{\partial h}{\partial a}+b\frac{\partial h}{\partial b} 
	\punto
\end{equation}

\section{Conclusions}
In this paper, we present an alternative parametric model for the definition of the functional shape HRF that requires less parameter compared to the standard canonical HRF, which potentially eases its application in practice. As future work, the advances and applicability of the proposed method will be tested on real fMRI data.

\newpage

%%% ┬┐ ┬ ┬┐ ┬  ┬┌─┐┌─┐┬─┐┌─┐┌─┐┬ ┬┬ ┬ %%%%%%%%
%%% ├┴┐│ ├┴┐│  ││ ││ ┬├┬┘├─┤├─┘├─┤└┬┘ %%%%%%%%
%%% ┴─┘┴ ┴─┘┴─┘┴└─┘└─┘┴└─┴ ┴┴  ┴ ┴ ┴  %%%%%%%%
 
\bibliographystyle{ieeetr}
\bibliography{MyBiblio}

\end{document}